\documentclass{pasa}

\def\aj{AJ}%
\def\apj{ApJ}%
\def\apjl{ApJ}%
\def\apjs{ApJS}%
\def\aap{A\&A}%
\title[Recalibrating the WISE W4 Filter]{Recalibrating the Wide-field Infrared Survey Explorer (WISE) W4 Filter}
\author[Brown et al.]{M. J. I. Brown$^{1,2,3}$, T. H. Jarrett$^4$, \and M. E. Cluver$^{5}$\\
\affil{$^1$School of Physics, Monash University, Clayton, Victoria 3800, Australia}%
\affil{$^2$Monash Centre for Astrophysics, Monash University, Clayton, Victoria, 3800, Australia}%
\affil{$^3$ARC Future Fellow}%
\affil{$^4$Department of Astronomy, University of Cape Town, Private Bag X3, Rondebosch 7701, South Africa}%
\affil{$^5$University of the Western Cape, Robert Sobukwe Road, Bellville, 7535, South Africa}}%
\jid{PASA}
\doi{10.1017/pas.\the\year.xxx}
\jyear{\the\year}

\usepackage[authoryear]{natbib}
\bibpunct{(}{)}{;}{a}{}{,}
\begin{document}%
\begin{abstract}
We present a revised effective wavelength and photometric calibration for the Wide-field Infrared Survey Explorer (WISE) $W4$ band, including tests of empirically motivated modifications to its pre-launch laboratory-measured relative system response curve. We derived these by comparing measured $W4$ photometry with photometry synthesised from spectra of galaxies and planetary nebulae. The difference between measured and synthesised photometry using the pre-launch laboratory-measured $W4$ relative system response can be as large as $0.3~{\rm mag}$ for galaxies and $1~{\rm mag}$ for planetary nebulae. We find the $W4$ effective wavelength should be revised upward by 3.3\%, from $22.1~{\rm \mu m}$ to $22.8~{\rm \mu m}$, and the $W4$ AB magnitude of Vega should be revised from $m_{W4}=6.59$ to $m_{W4}=6.66$.  In an attempt to reproduce the observed $W4$ photometry, we tested three modifications to the pre-launch laboratory-measured $W4$ relative system response curve, all of which have an effective wavelength of $22.8~{\rm \mu m}$. Of the three relative system response curve models tested, a model that matches the laboratory-measured relative system response curve, but has the wavelengths increased by 3.3\% (or $\simeq 0.73~{\rm \mu m}$) achieves reasonable agreement between the measured and synthesised photometry.
\end{abstract}
\begin{keywords}
infrared: general -- infrared: galaxies -- galaxies: photometry -- (ISM:) planetary nebulae: general -- space vehicles: instruments
\end{keywords}
\maketitle%
\section{Introduction}
\label{sec:intro}

The Wide-field Infrared Survey Explorer \citep[WISE; ][]{wri10} surveyed the entire sky in four passbands, which were intended to have effective wavelengths of  $3.4$, $4.6$, $12$ and $22~{\rm \mu m}$. The two reddest filters, $W3$ and $W4$, are useful for constraining the thermal emission from objects with temperatures of several hundred Kelvin \citep[e.g.,][]{mai11a,mai11b,wu12}. The $W4$ filter also provides a star formation rate indicator for nearby galaxies that should be comparable in accuracy to {\it Spitzer} $24~{\rm \mu m}$ \citep[e.g.,][]{ken09,jar13,clu14}. 

Unfortunately, the pre-launch laboratory-measured WISE $W4$ relative system response (RSR) curve does not match its on-sky performance \citep{wri10}. As WISE photometry was calibrated using A stars and K-M giants \citep{jar11}, by construction objects with spectral indices between 1 and 2 (where the spectral index is defined by $f_\nu \propto \nu^\alpha$) have accurate photometry. However, the AB  magnitudes and flux densities for star-forming galaxies and active galactic nuclei with spectral indices of $\alpha <0$ are systematically in error. 

\citet{wri10} found that the fluxes of star-forming galaxies (with spectra approximated by $f_\nu \propto ^{-2}$) were overestimated by 9\%. \citet{bro14} found the error is a smooth function of $22~{\rm \mu m}$ spectral index, with the error approaching 30\% for some ultra-luminous infrared galaxies (ULIRGS). \citet{wri10} concluded that the flux errors could be resolved by shifting the effective wavelength of the $W4$ filter redward by 2\% to 3\%.

The aim of this paper is to define an accurate effective wavelength, photometric calibration, and practical RSR model for the WISE $W4$ filter. For filters with effective wavelengths of $\sim 22~{\rm \mu m}$, the AB magnitudes of Vega were determined using a single-temperature Kurucz model for Vega \citep{kur91,coh92}, incorporating a 2.7\% increase in the fluxes as described by \citet{coh09}. This Vega basis has an overall systematic uncertainty of $\simeq 1.45\%$ \citep{coh92}. The resulting AB magnitude of Vega is $m_{22}=5.81$ in the {\it Spitzer} IRS Peak-Up Imager red channel  and $m_{24}=6.67$ in the {\it Spitzer} MIPS $24~{\rm \mu m}$ band. In general we use Vega-based magnitudes, with all exceptions being explicitly noted in the text.

\section{Effective Wavelength}

In this paper we define apparent magnitudes using
\begin{equation}
m = -2.5 {\rm log} \left[ \left( \int R(\nu) \frac{f_\nu(\nu)}{h\nu} d\nu \right) \times \left( \int R(\nu) \frac{g_\nu(\nu)}{h\nu} d\nu \right)^{-1}  \right]
\label{eq:mag}
\end{equation}
\citep[e.g.,][]{hog02} where $f_\nu$ is the SED of the source, $g_\nu$ is the SED of a $m=0$ source (either Vega or $f_\nu = 3631~{\rm Jy}$), $R_\nu$ is the filter response function (defined as electrons per incident photon) and $h\nu$ is the energy of a photon with frequency $\nu$.  Please note that these magnitudes are based on photon counts rather than fluxes, and do not simply correspond to monochromatic flux densities at the effective wavelengths of the relevant filters.

\begin{figure}
\resizebox{3.25in}{!}{\includegraphics{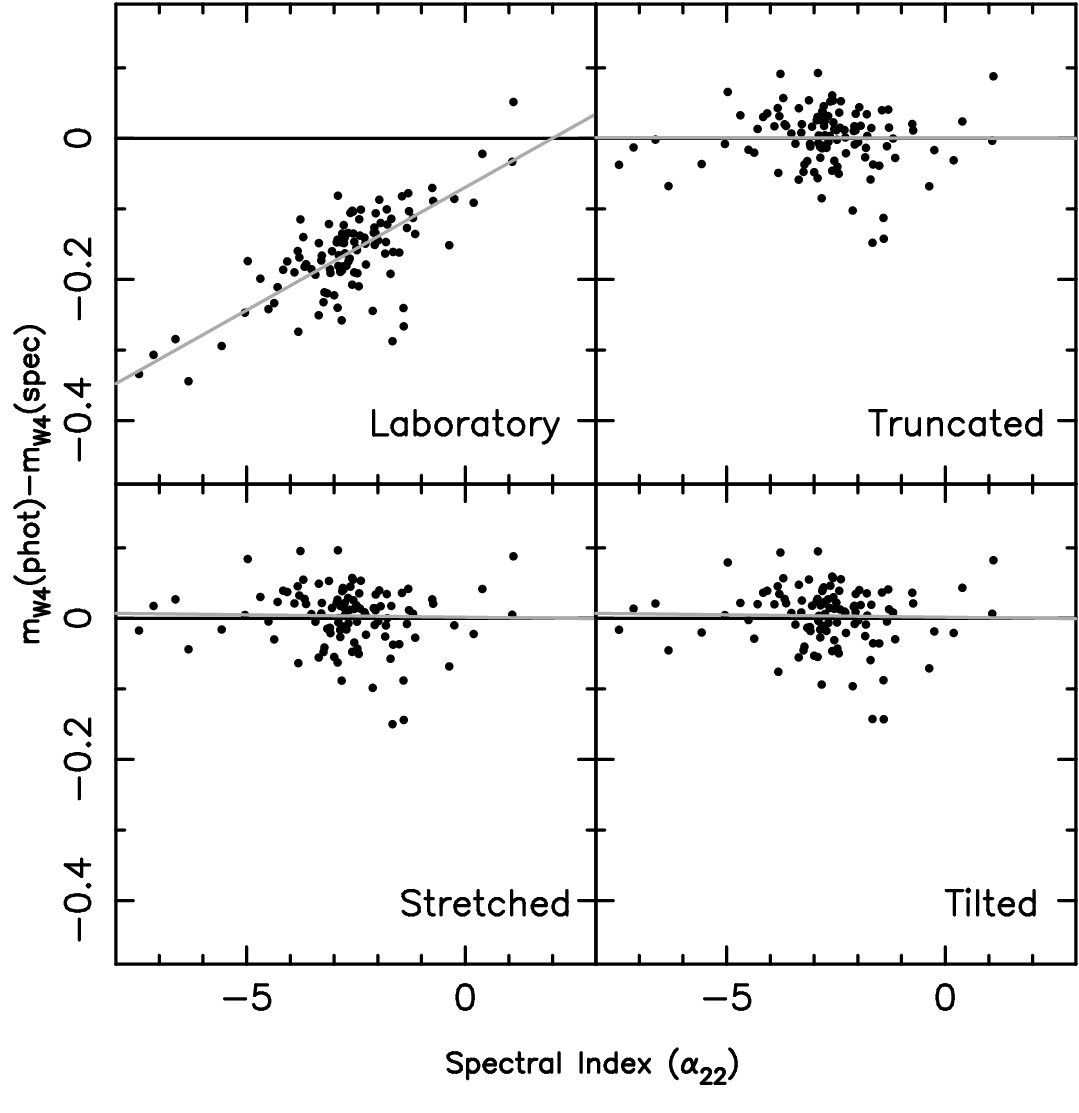}}
\caption{The difference between measured and synthesised $W4$ magnitudes for galaxies drawn from \citet{bro14}, plotted as a function of $\sim 22~{\rm \mu m}$ spectral index. The top-left panel shows that the pre-launch laboratory-measured WISE $W4$ RSR does not match the on-sky performance \citep{wri10,jar11}, so the measured $W4$ magnitudes are systematically too bright for galaxies with spectra that differ significantly from the Rayleigh-Jeans approximation. All three modified filter curves reduce the discrepancy between the measured the synthesised $W4$ magnitudes, with the stretched and tilted filter response curves performing best.
}
\label{fig:res}
\end{figure}

\citet{bro14} quantified the systematic error in the WISE $W4$ photometry by comparing measured photometry with synthetic photometry derived from galaxy spectral energy distributions. The \citet{bro14} sample spans a broad range of galaxy types, including early-type galaxies, late-type galaxies, luminous infrared galaxies, starbursts, Seyferts and blue compact dwarfs. The top-left panel of Figure~\ref{fig:res} shows the anomaly as a function of $\sim 22~{\rm \mu m}$ spectral index, $\alpha_{22}$. 

The residual is well approximated by 
\begin{equation}
\Delta m_{W4}  = (0.035 \pm 0.001) \times (\alpha_{22} - 2 )  
\label{eq:dm}
\end{equation}
with an RMS of 0.05 magnitudes \citep{bro14}.
This residual is much larger than the flux corrections discussed by \citet{wri10}, which were intended for derivations of monochromatic flux densities from $W4$ photometry. For a minority of objects with measured $\alpha_{22}$ or with an assumed $\alpha_{22}$, Equation~\ref{eq:dm} can be used to correct WISE $W4$ photometry so that it matches the pre-launch RSR curve \citep[e.g.,][]{bro14}. For the majority of objects that lack known $\alpha_{22}$ values, we need a revised effective wavelength and WISE $W4$ RSR model.  

We can use Equation~\ref{eq:dm} to derive the revised effective wavelength for the $W4$ filter. If the pre-launch effective wavelength and actual effective wavelengths correspond to frequencies of $\nu_0$ and $\nu_1$ respectively, then 
\begin{equation}
\Delta m_{W4} \simeq - 2.5 {\rm log} \left( \frac{f_\nu(\nu_1)}{g_\nu(\nu_1)} \right) + 2.5 {\rm log} \left( \frac{f_\nu(\nu_0)}{g_\nu(\nu_0)} \right)  
\end{equation}
\begin{equation}
\Delta m_{W4}  \simeq 2.5 {\rm log} \left( \frac{f_\nu(\nu_0)g_\nu(\nu_1)}{g_\nu(\nu_0) f_\nu(\nu_1) } \right)  
\end{equation}
If $g_\nu$ is a Rayleigh-Jeans spectrum and $f_\nu$ is a power-law with index $\alpha_{22}$ then
\begin{equation}
\Delta m_{W4}  \simeq  2.5 {\rm log} \left( \frac{f_\nu(\nu_0)g_\nu(\nu_0)\nu_1^{2}\nu_0^{-2}}{g_\nu(\nu_0)f_\nu(\nu_0) \nu_1^{\alpha}\nu_0^{-\alpha}} \right)  
\end{equation}
\begin{equation}
(0.035 \pm 0.001) \times (\alpha_{22} - 2 ) \simeq  2.5 {\rm log} \left( \frac{\nu_0^{(\alpha_{22}-2)}}{\nu_1^{(\alpha_{22}-2)}} \right)  
\end{equation}
\begin{equation}
(0.035 \pm 0.001) \simeq  2.5 {\rm log} \left( \frac{\nu_0}{\nu_1} \right)  
\label{eq:shift}
\end{equation}

Equation~\ref{eq:shift} shows the $W4$ effective wavelength should be revised upward by $3.3\pm0.1\%$, from $22.1~{\rm \mu m}$ \citep{wri10} to $22.8~{\rm \mu m}$. The $W4$ filter can be considered, to first order, a $23~{\rm \mu m}$ passband. The corresponding AB magnitude of Vega in the $W4$ band is revised from $m_{W4}=6.59$ to $m_{W4}=6.66$. When updating existing WISE catalogues to use the new effective wavelength, Vega-based magnitudes can remain unchanged in value, but AB magnitudes (and SED dependent monochromatic flux densities) must account for the revised AB magnitude of Vega. 

\section{RSR Models and Galaxies}

To compare observed and model spectral energy distributions with observed $W4$ photometry, a revised $W4$ RSR model is needed. We have tested the pre-launch laboratory-measured $W4$ RSR and three alternative $W4$ RSR models to determine which provides the best agreement between measured photometry and photometry synthesised from spectra. All three alternative $W4$ RSR models have an effective wavelength of $22.8~{\rm \mu m}$. 

\begin{figure}
\resizebox{4.0in}{!}{\includegraphics{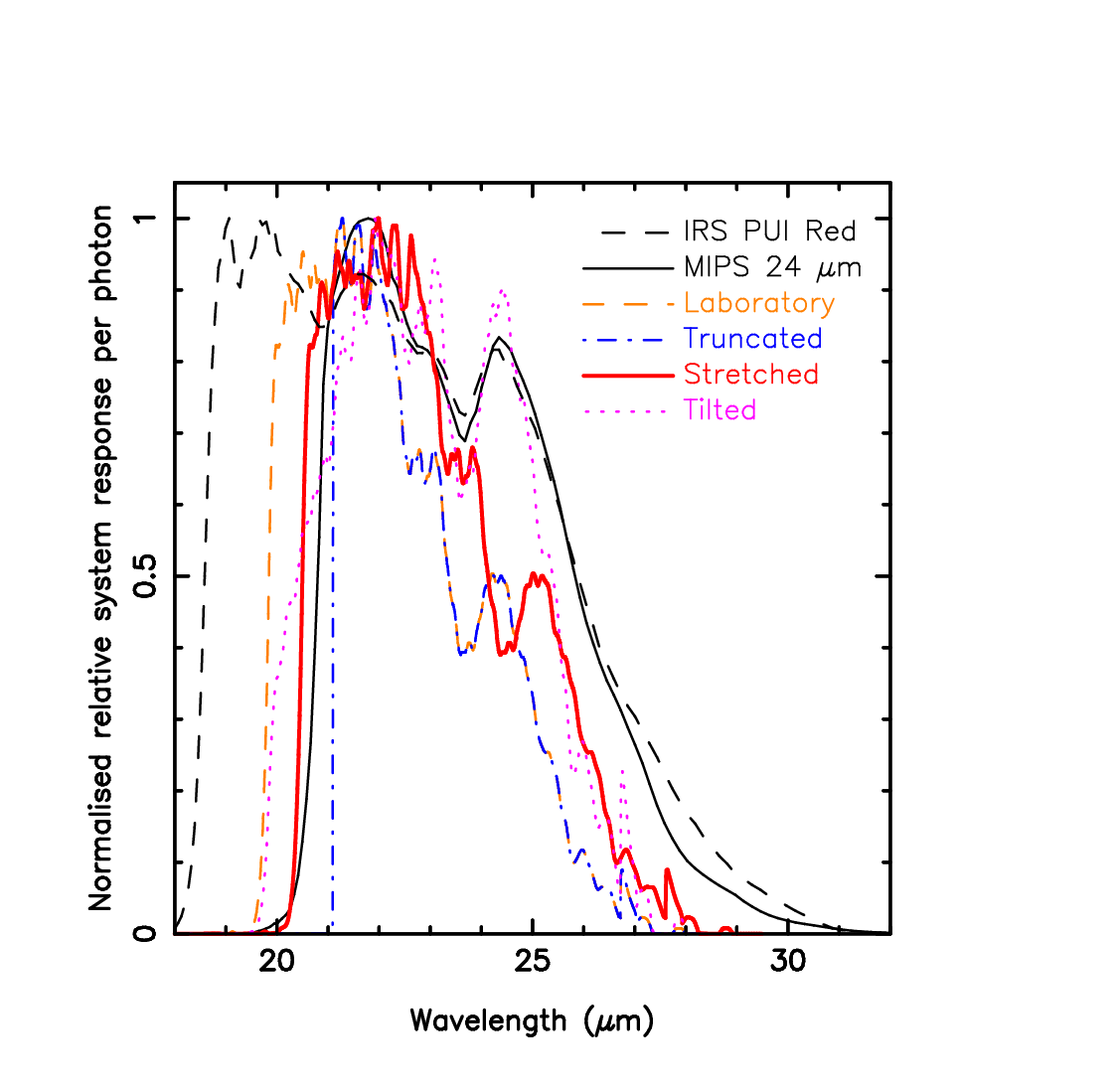}}
\caption{The laboratory-measured $W4$ RSR, the three modified $W4$ RSRs, the {\it Spitzer} IRS Peak-Up Imager red channel ($22.3~{\rm \mu m}$) RSR and the MIPS $24~{\rm \mu m}$ RSR. All of the RSRs are renormalised so the RSRs peak at $1.00$. 
}
\label{fig:filters}
\end{figure}

The truncated RSR model is identical to the laboratory-measured RSR except it has zero transmission below a wavelength of $21.1~{\rm \mu m}$. The stretched RSR is identical to the laboratory-measured RSR except all the wavelengths have been revised upward by 3.3\% (i.e., $\Delta \lambda = 0.033\lambda$). Please note that comparisons of measured and synthesised photometry using the stretched RSR and a modified laboratory-measured RSR with the wavelengths increased by $0.73~{\rm \mu m}$ (i.e., $\Delta \lambda = 0.73~{\rm \mu m}$) were equivalent within our uncertainties. The tilted RSR is the product of the laboratory-measured RSR multiplied by a function that increases linearly from 0 to 1 between $18.6~{\rm \mu m}$ and $28~{\rm \mu m}$ wavelength. All of the RSRs used in this paper are plotted in Figure~\ref{fig:filters}.

We used the \citet{bro14} {\it Spitzer} IRS-LL spectra and models of galaxies to generate synthetic photometry for the laboratory-measured $W4$ RSR and the three alternative $W4$ RSR models, and then compared this to the measured matched aperture photometry of \citet{bro14}. As we illustrate in Figure~\ref{fig:res}, all three modified $W4$ RSRs produce excellent agreement between measured photometry and photometry synthesised from spectra for galaxies, with no significant residual as a function spectral index.

\section{RSR Models and Planetary Nebulae}

To further discriminate between the RSR models, we have measured the mid-infrared colours of bright planetary nebulae with WISE $W4$ and {\it Spitzer} MIPS $24~{\rm \mu m}$ filters, and compared these with colours synthesizsd from {\it Spitzer} IRS-LL spectroscopy. While galaxy spectra are dominated by a (relatively) featureless continuum near $23~{\rm \mu m}$, planetary nebulae often feature a prominent ${\rm [OIV]}$ emission line at $25.9~{\rm \mu m}$. Although the planetary nebulae spectra in the $W4$ wavelength range are dominated by a single emission line, the magnitudes still depend on the shape of the entire $W4$ RSR as they are normalised with a model Vega spectrum. As the ${\rm [OIV]}$ emission line is far from the effective wavelength of the $W4$ filter, it is non-trivial to convert AB magnitudes (defined by Equation~\ref{eq:mag}) into monochromatic flux densities. 

Planetary nebulae from the NGC and IC catalogues were selected from \citet{koh01}, and these were then cross matched with data in the {\it Spitzer Heritage Archive}\footnote{http://sha.ipac.caltech.edu/} and the Cornell Atlas of Spitzer/IRS Sources \citep[CASSIS; ][]{leb11}. {\it Spitzer} IRS-LL spectra of planetary nebulae are typically ``stare mode'' spectra rather than ``spectral maps'', so we could not easily match our photometric aperture with the spectroscopic extraction aperture. However, the stare mode spectra were sufficient to check that the planetary nebulae selected had weak continua and strong ${\rm [OIV]}$ emission. We also attempted to use supernova remnants, but found most of the objects with IRS-LL spectra had relatively strong continua and weak ${\rm [OIV]}$ emission, so we excluded them from our analysis.

\begin{figure}
\resizebox{3.25in}{!}{\includegraphics{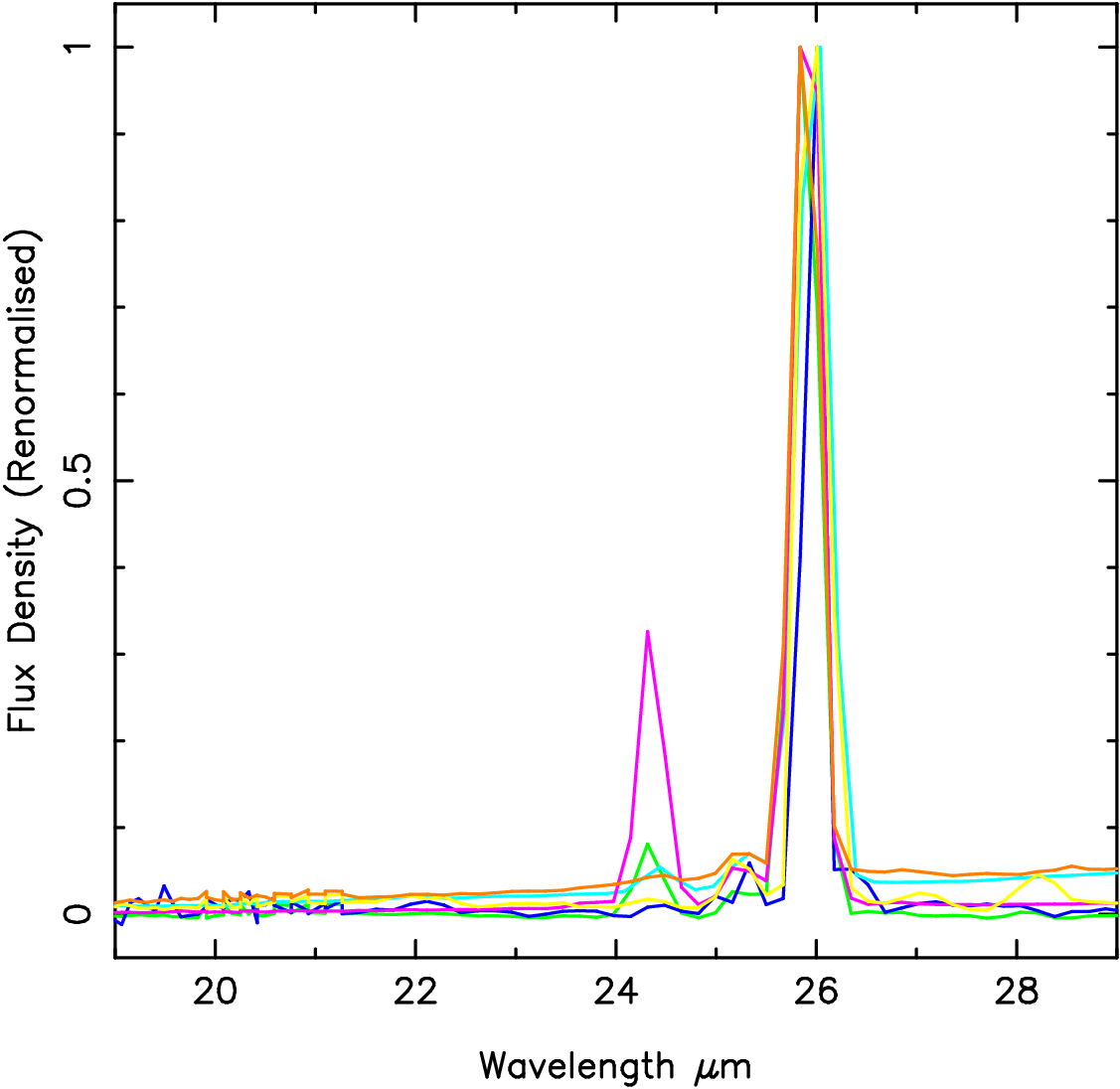}}
\caption{The {\it Spitzer} IRS-LL planetary nebula spectra used for testing the shape of the $W4$ RSR. To aid the comparison of the spectra, they have all been renormalised using the peak of the ${\rm [OIV]}$ emission line at $25.9~{\rm \mu m}$. All of the planetary nebulae have weak continuum emission and strong ${\rm [OIV]}$ emission, and some of the nebulae feature the ${\rm [NeV]}$ emission line at $24.3~{\rm \mu m}$.
}
\label{fig:spectra}
\end{figure}

The final sample of planetary nebulae was NGC~246, NGC~3587, NGC~6720, NGC~6852, NGC~6853 and NGC~7293, and the renormalised spectra are plotted in Figure~\ref{fig:spectra}. The photometry of the planetary nebulae was measured using large rectangular apertures as in \citet{bro14}, and this photometry is provided in Table~\ref{table:phottab}. We have not removed stellar contamination  from the planetary nebula photometry, but stars only significant alter the photometry well short-ward of the $W4$ filter.  As all of the planetary nebulae are very bright, we expect systematic errors (e.g., calibration errors) to dominate over photon Poisson noise. 

\begin{figure}
\resizebox{3.25in}{!}{\includegraphics{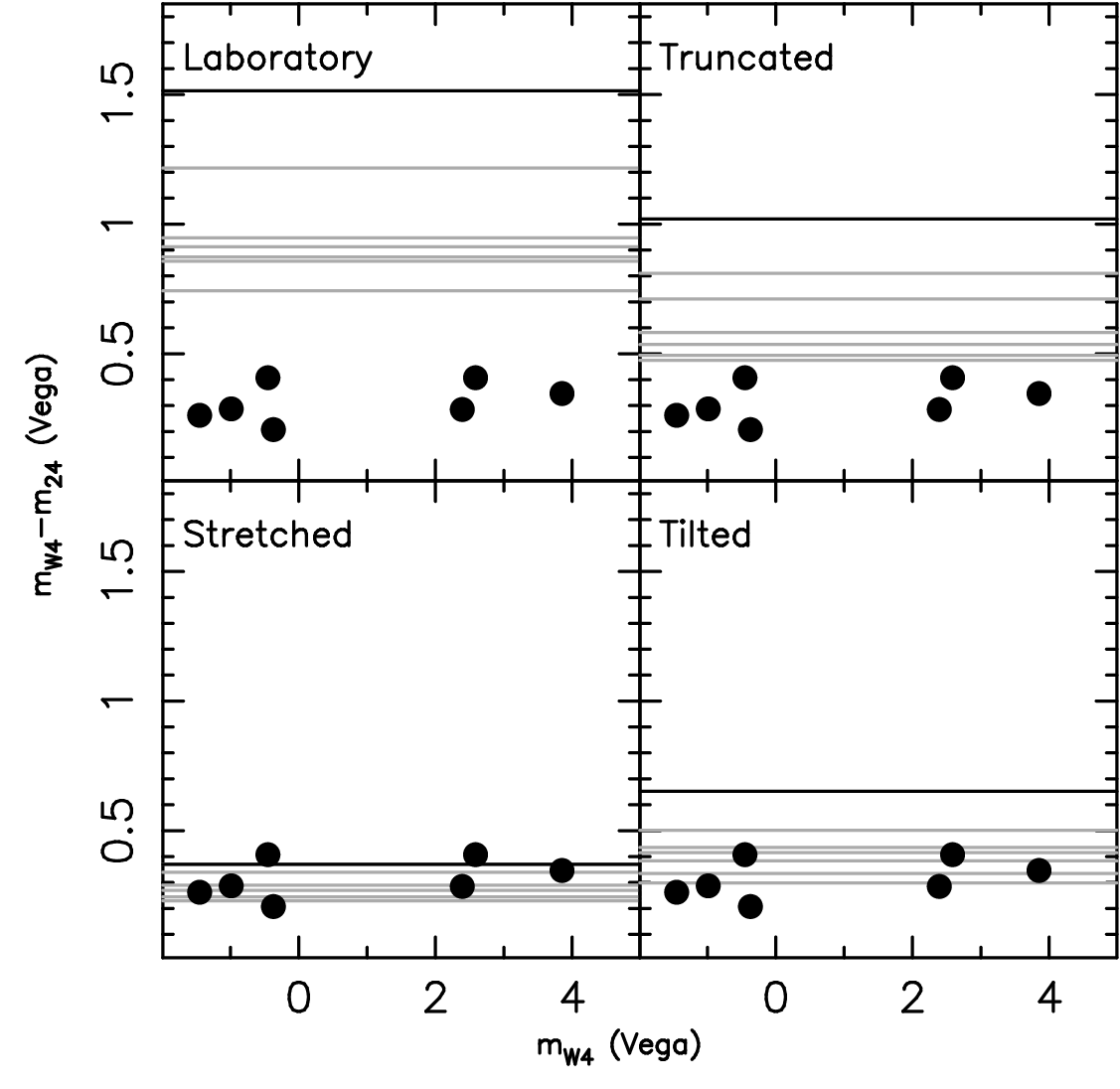}}
\caption{The measured $m_{W4}-m_{24}$ colours of planetary nebula and colours synthesised from spectra. The black solid line denotes a spectrum with a single $25.9~{\rm \mu m}$ emission line while the grey lines are synthesised from {\it Spitzer} IRS-LL spectra of planetary nebulae. Of the RSRs tested, the stretched filter response curve provides the best agreement between the measured photometry and photometry synthesised from spectra. 
}
\label{fig:planetary}
\end{figure}

In Figure~\ref{fig:planetary} we compare the measured $m_{W4}-m_{24}$ colours of planetary nebula with colours synthesised from spectra. There are gross offsets between the measured colours and those synthesised from the laboratory RSR and the truncated RSR. Smaller offsets are seen between measured colours and those synthesised from the tilted RSR, while there is good agreement between the measured and synthesised photometry for stretched RSR. Both the tilted and stretched RSRs effectively increase transmission at the red end of the filter (i.e., they are consistent with a ``red leak''). 

The {\it Spitzer} IRS Peak-Up Imager red channel ($22.3~{\rm \mu m}$) allows us to cross check the measured and synthesised colours of planetary nebulae without relying on calibration of the MIPS $24~{\rm \mu m}$ band, although we can undertake this test for just one object in our sample. The measured colour of NGC 6852 is $m_{22}-m_{W4}=0.86$, while a single $25.9~{\rm \mu m}$ emission line produces $m_{22}-m_{W4}$ colours of $-0.22$, $0.28$, $0.93$ and $0.65$ for the laboratory-measured, truncated, stretched and tilted RSRs (respectively). This is consistent with our tests using $m_{W4}-m_{24}$ colours, where the best agreement between measured and synthesised colours was for the stretched RSR.

We caution that only three RSR models are presented in this paper, and our stretched RSR is unlikely to be a unique solution for resolving the discrepancy between measured and synthesised $W4$ photometry. We also caution that while the stretched RSR provides the best agreement between the measured and synthesised $W4$ photometry, an error in transmission or quantum efficiency is far more likely than a wavelength error. The stretched filter curve should thus be considered a useful empirical model for comparing $W4$ photometry with spectra, rather than a unique physical description of the $W4$ RSR. 

\section{Summary}

We present a revised effective wavelength, photometric calibration, and RSR models for the WISE $W4$ filter, derived using comparisons of measured photometry and photometry synthesised from galaxy and planetary nebula spectra. The on-sky performance of the WISE $W4$ RSR does not match pre-launch laboratory measurements, resulting in large flux density errors for star-forming galaxies, active galactic nuclei and planetary nebulae.

The offset between the measured $W4$ photometry and photometry synthesised from spectra is a function of $22~{\rm \mu m}$ spectral index. When spectral index information is available, Equation~{\ref{eq:dm}, can be used to adjust measured photometry so it is on the system defined by the pre-launch laboratory-measured RSR. 

When spectral index information is not available, a revised effective wavelength must be used (although Vega-based magnitudes remain unchanged). We find the effective wavelength of the $W4$ filter should be revised upward by 3.3\%, from $22.1~{\rm \mu m}$ to $22.8~{\rm \mu m}$. The corresponding AB magnitude of Vega in the $W4$ band should be revised from $m_{W4}=6.59$ to $m_{W4}=6.66$. The $W4$ filter can be considered, to first order, a $23~{\rm \mu m}$ passband.

To compare spectral energy distributions with observed $W4$ photometry, a revised $W4$ RSR model is needed. Using galaxy spectra, we could not easily distinguish between different models of the $W4$ RSR that all have an effective wavelength of $22.8~{\rm \mu m}$. The prominent $25.9~{\rm \mu m}$ ${\rm [OIV]}$ emission line in planetary nebula spectra allowed us to distinguish between three possible $W4$ RSR models. Of the three RSR models tested, we find that a RSR that matches the shape laboratory-measured curve, but with the wavelengths increased by 3.3\% (or increased by $\simeq 0.73~{\rm \mu m}$) adequately reconciles measured photometry and photometry synthesised from spectra.

\begin{acknowledgements}
MB acknowledges financial support from the Australian Research Council (FT100100280) and the Monash Research Accelerator Program (MRA). TJ and MC would like to acknowledge the support of the South African National Research Foundation and Department of Science and Technology.  This publication makes use of data products from the Wide-field Infrared Survey Explorer, which is a joint project of the University of California, Los Angeles, and the Jet Propulsion Laboratory/California Institute of Technology, funded by the National Aeronautics and Space Administration. This work is based in part on observations made with the {\it Spitzer} Space Telescope, obtained from the NASA/ IPAC Infrared Science Archive, both of which are operated by the Jet Propulsion Laboratory, California Institute of Technology under a contract with the National Aeronautics and Space Administration. The Cornell Atlas of Spitzer/IRS Sources (CASSIS) is a product of the Infrared Science Center at Cornell University, supported by NASA and JPL.
\end{acknowledgements}

\begin{table*}
\caption{Planetary nebula photometry\label{table:phottab}}
\begin{center}
\begin{tabular*}{\textwidth}{@{}c\x c\x c\x c\x c\x c\x c\x c\x c\x c\x c\x c@{}}
\hline \hline
Name & J2000 Coordinates & Photometric Aperture & $m_{W1}$$^{\rm a}$ & $m_{W2}$ & $m_{W3}$ & $m_{22}$ & $m_{W4}$ & $m_{24}$ \\
\hline
NGC 246    & 00:47:04 -11:52:14.1   & $300^{\prime\prime} \times 300^{\prime\prime}$     & 7.99   & 7.84   & 3.68   & -         & -0.96  & -1.28 \\
NGC 3587  & 11:14:47 +55:01:06.6  & $100^{\prime\prime} \times 100^{\prime\prime}$     & 10.68 & 10.37 & 5.83   &  -        & 3.88   &  3.51 \\
NGC 6720  & 18:53:34 +33:01:48.2  & $120^{\prime\prime} \times 120^{\prime\prime}$     & 7.84   & 7.14   & 2.62   & -         & -0.34  & -0.58 \\
NGC 6852  & 20:00:39 +01:43:43.5  & $45^{\prime\prime} \times 45^{\prime\prime}$         & 14.23  & 10.65  & 6.40 &  3.45  & 2.59   & 2.18 \\
NGC 6853  & 19:59:37 +22:43:26.4  & $480^{\prime\prime} \times 480^{\prime\prime}$     & 5.00   & 4.43   & 1.40   & -         & -1.42  & -1.71 \\
NGC 7293  &  22:29:39 -20:50:09.7  & $1200^{\prime\prime} \times 1200^{\prime\prime}$ & 5.13   & 4.55   & 0.95   & -         & -0.42  & -0.86 \\
\hline
\end{tabular*}
\end{center}
\tabnote{$^a$Foreground and background stars were not subtracted from the images, and can contribute significantly to the $W1$ and $W2$ photometry.}
\end{table*}
 

\end{document}